\documentclass[%
 amsmath,amssymb,
 aps,
]{revtex4-1}

\usepackage{graphicx}
\usepackage{dcolumn}
\usepackage{bm}

\begin{document}

\title{The topological order of the space\footnote{Essay written for the Gravity Research Foundation 2021 Awards for Essays on Gravitation.}}
\author{Jingbo Wang}
\email{shuijing@mail.bnu.edu.cn}
\affiliation{Institute for Gravitation and Astrophysics, College of Physics and Electronic Engineering, Xinyang Normal University, Xinyang, 464000, P. R. China}
 \date{\today}
\begin{abstract}
Topological order is a new type order that beyond Landau's symmetry breaking theory. The topological entanglement entropy provides a universal quantum number to characterize the topological order in a system. The topological entanglement entropy of the BTZ black hole was calculated and found that it coincides with that for fractional quantum Hall state. So the BTZ black holes have the same topological order with the fractional quantum Hall state. We conjecture that black holes in higher dimensions also have topological orders. Next we want to study the topological order of ordinary spaces which can be described by spin network states in loop quantum gravity. We advise to bring in the methods and results in string-net condensation to loop quantum gravity to solve some difficult problems.
\end{abstract}

\bibliographystyle{unsrt}
\maketitle

In our world there are many different phases of matter. For a long time, it was believed that those phases can be described by Laudau's symmetry breaking theory. But in late 1980s, it was found that in chiral spin liquid there may exist a new kind of order, so-called topological order that beyond the usual symmetry description \cite{to1}. The term ``topological" was motivated by the fact that the low energy effective theory for the chiral spin liquid is a topological quantum field theory.  Since then, the study of topological phases of quantum matter slowly becomes more and more active, and then a main stream in condensed matter physics now. It was proposed that such phase of matter can be used to build powerful quantum computer \cite{qc1,qc2}.

The notion of topological order can be defined macroscopically by topological degeneracy and non-Abelian geometric phases of the ground states on spaces with non-trivial topology. Those properties are all robust against any small perturbations. It can also be defined microscopically as pattern of long range entanglement. Topological order can produce quasiparticles with fractional quantum numbers and fractional/Fermi statistics, robust gapless boundary modes and emergent gauge excitations. So they can provide an unification of gauge interaction and Fermi statistics \cite{wen3}.

A universal quantum number to characterize the topological order in a system can be provided by the topological entanglement entropy (TEE)\cite{tee1,tee2}. For a two-dimensional topological ordered system, the entanglement area law satisfy
\begin{equation}\label{1}
  S\sim \alpha L-\gamma,
\end{equation}
where $L$ is the length of the boundary, and $\alpha$ is a parameter that depend on the system, $\gamma>0$ is a constant term that indicates the existence of the topological order. This term is called the topological entanglement entropy. For example, the TEE for fractional quantum Hall state with filling factor $v=\frac{1}{m}$ takes the form
\begin{equation}\label{2a}
  \gamma=\frac{1}{2}\ln m.
\end{equation}

In the previous paper \cite{whcft1}, we started from the Chern-Simons theory formulation of gravity with negative cosmological constant, and applied some suitable boundary conditions, then found that for BTZ black holes the boundary degrees of freedom can be described by two chiral massless scalar fields $\Psi_L$ (for left-moving) and $\Psi_R$ (for right-moving). A chiral massless scalar field can also be used to describe the boundary modes of the fractional quantum Hall states, so from this point of view, the BTZ black hole can be considered as two copies of quantum Hall states, or a quantum spin Hall state. Quantizing those scalar fields gives the microscopic states of the BTZ black hole. The entropy for those scalar fields can be calculated to be
 \begin{equation}\label{2}\begin{split}
  S_R=\frac{\pi}{4}(r_++r_-)-\frac{1}{2}\ln \frac{r_++r_-}{8},\\
  S_L=\frac{\pi}{4}(r_+-r_-)-\frac{1}{2}\ln \frac{r_+-r_-}{8},\\
  S_{BH}=S_R+S_L=\frac{2\pi r_+}{4}-\frac{1}{2}\ln \frac{r^2_+-r^2_-}{64}.
\end{split}\end{equation}
The logarithmic term in (\ref{2}) for right sector is the same with the topological entanglement entropy for fractional quantum Hall state with filling factor $v_R=\frac{8}{r_++r_-}$, and the same is true for left sector. So we can say that the right (left) sector of the BTZ black hole has the same topological order with the fractional quantum Hall state with filling factor $v_R=\frac{8}{r_++r_-}$ ($v_L=\frac{8}{r_+-r_-}$).

For higher dimensional black holes, such as Kerr black holes in four dimension, the boundary modes can be described by massless scalar field, which is also the same as higher dimensional topological insulators \cite{wangbms4}. Just like the BTZ black hole, we can also identify the quantum states of the scalar field with the microstates of Kerr black holes. We conjecture that black holes in higher dimensions can also have topological orders.

Since black holes can have topological orders, a natural question is: can an ordinary space (without black hole) have topological order? A quantum description of space can be given by spin network states. They play important roles in loop quantum gravity, lattice gauge theory, topological quantum field theory and so on. So the question becomes if a general spin network state has topological order?

A physical picture for topological order is provided by string-net condensation in Levin-Wen model\cite{lw1}. The Levin-Wen model is believed to be a Hamiltonian version of the Turaev-Viro topological quantum field theory in three dimensional spacetime. This model can be generalized to higher dimension. The underlying mathematical framework is tensor category theory. They can realize all discrete gauge theories (in any dimension) and all doubled Chern-Simons theories (in $2 + 1$ dimensions). Besides, they can provide exactly soluble Hamiltonian and ground state wave functions for each of these phases. Actually spin networks are very similar to string-nets, so in Ref.\cite{bv1} the authors suggest to bring in the string-nets into loop quantum gravity to include the fermionic and bosonic excitations. Bringing in the string-nets into loop quantum gravity can also have other advantages:
\begin{itemize}
\item The Hamiltonian for the string-nets is exactly soluble. It takes the form

 \begin{equation}\label{3}
   H=-\sum_I Q_I-\sum_p B_p, \quad B_p=\sum_{s=0}^N a_s B_p^s,\quad a_s=\frac{d_s}{\sum_{i=0}^N d_i^2},
\end{equation}
where the sums rum over vertices $I$ and the plaquettes $p$ of the lattice. The first term $Q_I$ is an electric charge operator, and the second term $B_p$ is a linear combination of magnetic flux operators $B_p^s$.

A difficult problem in loop quantum gravity is that due to the complicate form of the Hamiltonian constraint, it is very difficult to find exact eigenstates. One can modify the Hamiltonian for the string-nets to get a new Hamiltonian which is exactly soluble and also in the same universal class with that in loop quantum gravity. This is just like the case in fractional quantum Hall effect. Due to the interaction between electrons the Hamiltonian is very complicate and the exact ground states is difficult to find. One can construct a toy Hamiltonian which gives the Laughlin states that lie in the same universal class with that in true fractional quantum Hall effect.
\item In topological phase of matter, a widely observed phenomenon is the boundary-bulk correspondence \cite{bounbulk1,bounbulk2}, which relates the topological structure of the bulk states to the presence of protected zero-energy boundary states. This boundary-bulk correspondence can be considered as a weak version of the holographic principle \cite{hp1,hp2} in gravity theory. Actually in Ref.\cite{hp1}, it was suggest that ``quantum gravity should be described entirely by a topological quantum field theory". And topological quantum field theory can describe the low energy behavior of topological phase of matter.
\item On the other hand, the studies in loop quantum gravity can also be applied to the researches in string-net condensation theory. A main problem in the unification through string-net condensation is how to construct the graviton \cite{gw1}. But in loop quantum gravity, there are some attempts to construct graviton form loop or spin network \cite{gw2,gw3,gw4}, which may be useful for string-net condensation.
    \end{itemize}

In summary, we can use the methods in string-net condensation to study the spin network states in loop quantum gravity. If space has topological order, it can support gauge field and Fermion field from simple bosonic qubits. It may solve the origin of elementary particles in the standard model, and resents an unification of matter and information \cite{wen3}.


\acknowledgments
 This work is supported by Nanhu Scholars Program for Young Scholars of XYNU.

\bibliography{essay2021}

\begin{thebibliography}{10}

\bibitem{to1}
X.~G. {Wen}.
\newblock {Topological Orders in Rigid States}.
\newblock {\em International Journal of Modern Physics B}, 4(2):239--271,
  January 1990.

\bibitem{qc1}
A.~Y. Kitaev.
\newblock {Fault tolerant quantum computation by anyons}.
\newblock {\em Annals Phys.}, 303:2--30, 2003.

\bibitem{qc2}
C.~Nayak, S.~H. Simon, A.~Stern, M.~Freedman, and S.~Das~S.
\newblock {Non-Abelian anyons and topological quantum computation}.
\newblock {\em Rev. Mod. Phys.}, 80:1083--1159, 2008.

\bibitem{wen3}
B.~Zeng, X.~Chen, D.-L. Zhou, and X.~G. Wen.
\newblock {\em {Quantum Information Meets Quantum Matter: From Quantum
  Entanglement to Topological Phases of Many-Body Systems}}.
\newblock Springer, New York, 2019.

\bibitem{tee1}
A.~Kitaev and J.~Preskill.
\newblock {Topological entanglement entropy}.
\newblock {\em Phys. Rev. Lett.}, 96:110404, 2006.

\bibitem{tee2}
M.~Levin and X.-G. Wen.
\newblock {Detecting Topological Order in a Ground State Wave Function}.
\newblock {\em Phys. Rev. Lett.}, 96:110405, 2006.

\bibitem{whcft1}
J.~Wang and C.-G. Huang.
\newblock {Conformal field theory on the horizon of a BTZ black hole}.
\newblock {\em Chin. Phys.}, C42(12):123110, 2018.

\bibitem{wangbms4}
J.~Wang.
\newblock {Microscopic states of Kerr black holes from boundary-bulk
  correspondence}.
\newblock {\em Chin. Phys.}, C45(1):015107, 2021.

\bibitem{lw1}
M.~A. Levin and X.-G. Wen.
\newblock String-net condensation:a physical mechanism for topological phases.
\newblock {\em Physical Review B}, 71(4), Jan 2005.

\bibitem{bv1}
S.~Bilson-Thompson and D.~Vaid.
\newblock {LQG for the Bewildered}.
\newblock 2 2014.

\bibitem{bounbulk1}
M.~Z. Hasan and C.~L. Kane.
\newblock {Topological Insulators}.
\newblock {\em Rev. Mod. Phys.}, 82:3045, 2010.

\bibitem{bounbulk2}
X.~L. Qi and S.~C. Zhang.
\newblock {Topological insulators and superconductors}.
\newblock {\em Rev. Mod. Phys.}, 83(4):1057--1110, 2011.

\bibitem{hp1}
G.~'t~Hooft.
\newblock {Dimensional reduction in quantum gravity}.
\newblock {\em Conf. Proc. C}, 930308:284--296, 1993.

\bibitem{hp2}
L.~Susskind.
\newblock {The World as a hologram}.
\newblock {\em J. Math. Phys.}, 36:6377--6396, 1995.

\bibitem{gw1}
Z.-C. Gu and X.-G. Wen.
\newblock {Emergence of helicity +- 2 modes (gravitons) from qbit models}.
\newblock {\em Nucl. Phys. B}, 863:90--129, 2012.

\bibitem{gw2}
Z.~Joost.
\newblock Gravitons in loop quantum gravity.
\newblock {\em Nuclear Physics B}, 378(1):288--308, 1992.

\bibitem{gw3}
A.~Ashtekar, C.~Rovelli, and L.~Smolin.
\newblock {Gravitons and loops}.
\newblock {\em Phys. Rev. D}, 44:1740--1755, 1991.

\bibitem{gw4}
E.~Bianchi, L.~Modesto, C.~Rovelli, and S.~Speziale.
\newblock {Graviton propagator in loop quantum gravity}.
\newblock {\em Class. Quant. Grav.}, 23:6989--7028, 2006.

\end{thebibliography}

\end{document}